\documentstyle[epsf,floats,prc,amsfonts,amsmath,prl,aps,epsfig,twocolumn]{revtex}

\newcommand{\lto}{\leftarrow}
\newcommand{\dto}{\downarrow}
\newcommand{\vs}{\vspace{0.2cm}}

\newcommand{\ket}[1]{\left|#1\right\rangle}
\newcommand{\bra}[1]{\left\langle#1\right|}

\newcommand{\calQ}{{\mathcal{Q}}}
\newcommand{\calO}{{\mathcal{O}}}
\newcommand{\calF}{{\mathcal{F}}}

\newcommand{\Z}{{\mathbb{Z}}}

\begin{document}

\draft  

\wideabs{ 

\title{Time Evolution of Two-Level Systems Driven by Periodic Fields}
 
\author{J. C. A. Barata and D. A. Cortez}

\address{   
    Instituto de F\'{\i}sica.
    Universidade de S\~ao Paulo\\
    Caixa Postal 66 318.
    05315 970 S\~ao Paulo SP. Brasil \\
    E-mail: jbarata{@}fma.if.usp.br and dacortez{@}fma.if.usp.br
}


\maketitle 

\begin{abstract}
  In this paper we study the time evolution of a class of two-level
  systems driven by periodic fields in terms of new convergent
  perturbative expansions for the associated propagator $U(t)$. The
  main virtue of these expansions is that they do not contain secular
  terms, leading to a very convenient method for quantitatively
  studying the long-time behaviour of that systems.  We present a
  complete description of an algorithm to numerically compute the
  perturbative expansions. In particular, we applied the algorithm to
  study the case of an ac-dc field (monochromatic interaction),
  exploring various situations and showing results on (time-dependent)
  observable quantities, like transition probabilities. For a simple
  ac field, we analised particular situations where an approximate
  effect of dynamical localisation is exhibited by the driven system.
  The accuracy of our calculations was tested measuring the unitarity
  of the propagator $U(t)$, resulting in very small deviations, even
  for very long times compared to the cycle of the driving field.
\end{abstract}


}


\section{Introduction} \label{sec:intro}

Periodically (or more generically quasi-periodically) driven quantum
two-level systems are of basic importance in many physical
applications an have been widely studied since the pioneering works of
Rabi \cite{Rabi}, of Bloch and Siegert~\cite{BlochSiegert} and of
Autler and Townes~\cite{AutlerTownes} (see
also~\cite{Walter1} for more recent discussions
and \cite{sacchetti,qp,bw} for other general references on the
subject). The time evolution of such systems is governed by the
Schr\"odinger equation (we henceforth adopt $\hbar = 1$)
\begin{equation} \label{eq:S} 
i \frac{d}{dt} \ket{\Psi} = H(t) \ket{\Psi} \, ,
\end{equation} 
where $\ket{\Psi} = \ket{\Psi(t)} = \left( \psi_{+}(t) \atop
  \psi_{-}(t) \right)$ and $H(t)$ is the Hamiltonian of the system.
We may consider, for instance, a spin-1/2 system under the influence
of a time-dependent (periodic) magnetic field $\vec{B}(t)$.  In this
situation, $H(t)$ takes the usual form $H(t) = -\frac{1}{2}\vec{B}(t)
\cdot \vec{\sigma}$, where $\vec{\sigma} = (\sigma_1, \sigma_2,
\sigma_3)$ are the Pauli matrices. The interest in the solution of
(\ref{eq:S}) in this case is not restricted to the investigation of
the quantum system. As first pointed by Feynman, Vernon and
Hellwarth~\cite{Feynman} (see also the discussion in~\cite{Bagrov}),
the quantum system is equivalent to the classical Hamiltonian system
describing a classical gyromagnet precessing in a magnetic field:
$\frac{d}{dt} \vec{\mathcal{S}} = - \vec{B}(t) \times
\vec{\mathcal{S}}$, where $\vec{\mathcal{S}}$ is a (three dimensional)
unit vector. In fact, the methods described below can be directly
applied to the analysis of this system as well, since the components
of $\vec{\mathcal{S}}$ can be written in terms of the components $
\psi_{\pm}(t)$ of $\ket{\Psi}$~\cite{Feynman,Bagrov}.

The evolution of the systems governed by (\ref{eq:S}) is determined by
the time evolution operator $U(t, s)$ (also known as the propagator)
which connects the state $\ket{\Psi(s)}$ at time $s$ to the state
$\ket{\Psi(t)}$ at time $t$:
$
\ket{\Psi(t)} = U(t, s) \ket{\Psi(s)} \,  
$.
Defining $U(t) = U(t, 0)$ one has $U(t, s)=U(t)U(s)^\dagger$.  For a
time-dependent Hamiltonian the propagator $U(t)$ can be computed via
the {\it Dyson expansion:}
\begin{equation} \label{eq:Dyson}
U(t) = \hat{1} + \sum_{n=1}^{\infty} (-i)^n \int_0^t H(t_1) \, dt_1 \cdots 
\int_0^{t_{n-1}} H(t_n) \, dt_n \, .
\end{equation} 
Although (\ref{eq:Dyson}) gives a straightforward manner to compute
$U(t)$, the series in the r.h.s. is not generally uniformly convergent
in time.  For practical purposes this gives rise to difficulties when
one is interested in the large-time behaviour of the system. For
instance, if one considers a periodic Hamiltonian of the form $H(t) =
\sum_{m} H_m e^{i m \omega t}$, with $H_0 \neq 0$, two successive
integrations in (\ref{eq:Dyson}) would produce a linear term in~$t$.
Higher order terms in~$t$ would appear with further integrations.
These polynomial terms are known as {\it secular terms} and they
plague the expansion of $U(t)$ in such a way that its uniform
convergence is spoiled.

Of particular interest is the situation where the Schr\"odinger equation 
(\ref{eq:S}) takes the form
\begin{equation} \label{eq:H1} 
i \frac{d}{dt} \ket{\Psi} = H_1(t) \ket{\Psi} \, ,
\quad
\mbox{with}
\quad
H_1(t) := \epsilon \sigma_3 - f(t) \sigma_1 \, ,
\end{equation} 
where $f(t)$ is a periodic function of time $t$ and $\epsilon$ is a real
constant. By a rotation of $\pi/2$ around the 2-axis, we get the equivalent
system
\begin{equation} \label{eq:H2} 
i \frac{d}{dt} \ket{\Phi} = H_2(t) \ket{\Phi} \, ,
\quad
\mbox{with}
\quad
H_2(t) := \epsilon \sigma_1 + f(t) \sigma_3 \, ,
\end{equation} 
where $\ket{\Phi} = \ket{\Phi(t)} = \exp(-i\pi\sigma_2/4)\ket{\Psi(t)}$.
The Hamiltonian in (\ref{eq:H1}) can be interpreted as describing a system
with an unperturbed diagonal Hamiltonian $H_0 := \epsilon \sigma_3$, 
subjected to a (periodic) time-dependent perturbation $H_I(t) := -f(t)
\sigma_1$, inducing a time-depending transition between the unperturbed
eigenstates of $H_0$. Of course, (\ref{eq:H1}) can be also interpreted as a
spin-1/2 system under a magnetic field 
$\vec{B}(t) = (2f(t), 0, -2\epsilon)$. 

In~\cite{qp,p,dacjcab1} we investigated the system described by
(\ref{eq:H1}) or (\ref{eq:H2}) in the situation where $f$ is a
periodic or quasi-periodic function of time and $\epsilon$ is
``small'' and a special perturbative expansion (power series in
$\epsilon$) was developed, whose main virtue is to be free of secular
terms. The algorithm employed involves an inductive
``renormalisation'' of a sort of effective field introduced through an
exponential Ansatz (the function $g$, to be introduced below).
Moreover, in the periodic case, we have been able in~\cite{p,dacjcab1}
to prove rigorously the uniform convergence of our expansions for all $t$, for
$\epsilon$ sufficiently small, provided the function $f$ satisfies the
special conditions I, II and III we describe below.

A feature of our method is the fact that we are able to present the
complete (and convergent) $\epsilon$-expansion for the {\it secular
  frequency} $\Omega$ (also known as {\it Rabi frequency}) associated
to the solution of (\ref{eq:H1})-(\ref{eq:H2}). This is particularly
important for the qualitative study of the large-time behaviour of
that solutions.  Another relevant conclusion is that, under conditions
I-III, our method provides a complete perturbative solution of
(\ref{eq:H1})-(\ref{eq:H2}) for the monochromatic field (ad-dc field),
except perhaps for spurious situations \cite{dacjcab1}. 

Since our series are uniformly convergent in time, it is possible to
use them to study the long-time behaviour of observable quantities
in a controlled way.  It turns out that our expansions are also very
practical for numerical calculations, providing very small errors even
with few terms.

In this paper we describe the algorithm employed in the numerical
computation of our perturbative solution and show the results obtained
in the particular case of the monochromatic interactions. We begin in
Section~\ref{sec:method} by given a quick review of the perturbative
method developed in~\cite{qp,p,dacjcab1}.  Section~\ref{sec:numerical}
presents a description of the numerical algorithm developed to compute
the expansions. Next, in Section~\ref{sec:results}, we show and
discuss the results obtained in the case of monochromatic
interactions.  Finally, in Section~\ref{sec:conclusions}, we draw some
conclusions and final remarks.


\section{Description of the Method} \label{sec:method}

In~\cite{qp}, it was shown that the propagator $U(t)$ associated with the
system (\ref{eq:H2}) can be written as
\begin{equation} \label{eq:prop}
U(t) = \left(
\begin{array}{cc}
  R(t) (1 + i g_0 S(t)) & -i \epsilon R(t) S(t) \vspace{0.2cm}\\
  -i \epsilon \overline{R(t)} \overline{S(t)} & \overline{R(t)} (1 - i
  \overline{g_0} \overline{S(t)})
\end{array}
\right) \, ,
\end{equation}
where 
\begin{equation} \label{eq:R}
R(t) := \exp\left(-i \int_0^t (f(\tau) + g(\tau)) \, d\tau\right)
\end{equation} 
and
\begin{equation} \label{eq:St}
S(t) := \int_0^t R(\tau)^{-2} \, d\tau \, ,
\end{equation} 
where $g$ is a particular solution of the generalised Riccati equation
\begin{equation} \label{eq:Riccati}
g(t)'- i g(t)^2 - 2 i f(t) g(t) + i \epsilon^2 = 0 \, ,
\end{equation} 
and $g_0 \equiv g(0)$. To solve (\ref{eq:Riccati}) we pose 
\begin{equation} \label{eq:Ansatz}
g(t) = \sum_{n = 1}^{\infty} G^{(n)}(t) \epsilon^n \, ,
\end{equation} 
where $G^{(n)}(t) := q(t) c_n(t)$, with
\begin{equation}
q(t) := \exp\left(i \int_0^t f(\tau) \, d\tau\right) \, . \nonumber
\end{equation}
Inserting (\ref{eq:Ansatz}) into (\ref{eq:Riccati}) yields a sequence of
recursive equations for the coefficients $c_n$, whose solutions are
\begin{eqnarray}
c_1(t) &=& \alpha_1 q(t) \, , \label{eq:c1} \\
c_2(t) &=& q(t)\left[i \int_0^t (\alpha_1^2 q(\tau)^2 - q(\tau)^{-2}) \,
d\tau + \alpha_2 \right] \, , \label{eq:c2} \\
c_n(t) &=& q(t) \left[i \left( \int_0^t \sum_{p=1}^{n-1} c_p(\tau)
c_{n-p}(\tau) \, d\tau \right) + \alpha_n \right] \, , \label{eq:cn}
\end{eqnarray}
for $n \geq 3$, where the $\alpha_n$ are arbitrary integration constants.
Our strategy consists in fixing these constants inductively to cancel the
secular terms. For instance, to cancel the secular term in $c_2$, the
integrand in (\ref{eq:c2}) cannot contain a constant term. Defining the mean
value of a quasi-periodic function $h$ by 
\begin{equation}
M(h) := \lim_{T \to \infty} \frac{1}{2T} \int_{-T}^T h(t) \, dt \, ,
\nonumber
\end{equation}
we must require that $M(\alpha_1^2 q^2 - q^{-2}) = 0$. This leads to
$\alpha_1^2 = \overline{M(q^2)} / M(q^2)$, provided $M(q^2) \neq
0$. We remark that the mean value of a quasi-periodic function $h$ equals 
its zero order Fourier coefficient. In~\cite{qp,p,dacjcab1} we identified 
three classes of quasi-periodic functions $f$ for which this procedure of 
elimination of secular terms can be applied to all orders (i.e., to all 
functions $c_n$). These classes are defined by the following conditions: 
\begin{eqnarray}
{\rm I.} &\;\;&  M(\calQ_0) \neq 0 \, , \nonumber \\
{\rm II.} &\;\;& M(\calQ_0) =    0 \;\; \mbox{but} \;\; 
M(\calQ_1) \neq 0 \, , \nonumber \\
{\rm III.} &\;\;& M(\calQ_0) = M(\calQ_1) = 0 \;\; \mbox{but} \;\; 
M(\calQ_3) \neq 0 \, , \nonumber
\end{eqnarray}
where $\calQ_0(t) := q(t)^2$, 
\begin{eqnarray}
\calQ_1(t) &:=& \calQ_0(t) \int_0^t \left( \calQ_0(\tau)^{-1} - 
M\left(\calQ_0^{-1}\right) \right) \, d\tau \, , 
\label{eq:Q1} \\ 
\calQ_3(t) &:=& \calQ_0(t) \int_0^t (\calQ_1(\tau) - M(\calQ_1))
\, d\tau \, . \label{eq:Q3}
\end{eqnarray}

Under condition II, the Ansatz (\ref{eq:Ansatz}) has to be slightly
modified to
\begin{equation} \label{eq:AnsatzII}
g(t) = \sum_{n = 1}^{\infty} \widetilde{G}^{(n)}(t) \lambda^n \, ,
\quad \text{with} \quad \lambda := \epsilon^2
\end{equation}
and $\widetilde{G}^{(n)}(t) := q(t) e_n(t)$. The solutions for $e_n$ are
\begin{eqnarray}
e_1(t) &=& q(t) \left( -i \int_0^t q(\tau)^{-2} \, d\tau + \beta_1\right)
\, , \label{eq:e1} \\
e_n(t) &=& q(t) \left[i \left( \int_0^t \sum_{p=1}^{n-1} e_p(\tau)
e_{n-p}(\tau) \, d\tau \right) + \beta_n \right] \, , \label{eq:en} 
\end{eqnarray} 
for $n \geq 2$. Under condition III, we found in~\cite{dacjcab1}  
the same solution for $g$ as in I, namely (\ref{eq:Ansatz}). 
The difference between the solutions obtained in I and III lie in the  
set of constants $\alpha_n$, which is conventionally fixed to remove the
secular terms in each case.

In~\cite{p} we have shown that the $\epsilon$-expansion (\ref{eq:Ansatz})
or (\ref{eq:AnsatzII}) converges uniformly for $|\epsilon|$ small and for 
all $t$, provided $f$ is periodic and satisfies I or II. 
In~\cite{dacjcab1} we extended this result for case III. 
As we will discuss soon, case III is particularly important
for the investigation of the dynamical localisation effect (also called,
less properly, coherent destruction of tunnelling) for monochromatic
interactions.

The secular frequency $\Omega$, in conditions I or III, is given 
by~\cite{p,dacjcab1,bw}
\begin{equation} \label{eq:secular}
\Omega = M(f) + M(g) = F_0 + \sum_{n=1}^\infty M\left(G^{(n)}\right)
\epsilon^n \, 
\end{equation}
and, in condition II, by
\begin{equation}
\Omega = F_0 + \sum_{n=1}^\infty M\left(\widetilde{G}^{(n)}\right)
\lambda^n \, . \nonumber
\end{equation}
Hence, with our previous definitions, for condition I, we have
\begin{eqnarray}
\Omega &=& F_0 + \epsilon \alpha_1 M(\calQ_0) \nonumber \\
&+& \epsilon^2 \left[ i \alpha_1^2 M(\calQ_2) - iM(\calQ_1) + \alpha_2
M(\calQ_0) \right] \nonumber \\
&+& \epsilon^3 \left[ 2 \alpha_1 M(\calQ_3) + \alpha_3 M(\calQ_0) \right] 
+ \calO(\epsilon^4) \, , \label{eq:OmegaI}
\end{eqnarray}
where 
\begin{equation}
\calQ_2(t) := \calQ_0(t) \int_0^t (\calQ_0(\tau) - M(\calQ_0)) \,
d\tau \, . \nonumber
\end{equation}
One easily sees, by computing the zero order Fourier coefficient of 
$\calQ_2$ (details in~\cite{dacjcab1}), that $M(\calQ_2) = 0$ 
whenever $M(\calQ_0) = 0$.  Hence, for condition II, we get 
\begin{equation} \label{eq:OmegaII}
\Omega = F_0 - i \epsilon^2 M(\calQ_1) + \calO(\epsilon^4) 
\end{equation}
and, for condition III,
\begin{equation} \label{eq:OmegaIII}
\Omega = F_0 + 2 \alpha_1 \epsilon^3 M(\calQ_3)
+ \calO(\epsilon^4) \, .
\end{equation}


\section{Numerical Computation of the Solution} \label{sec:numerical}

We present here a step-by-step algorithm which enables us to compute the
propagator (\ref{eq:prop}) solely from the Fourier coefficients of the
(periodic) interaction $f$. This algorithm is based on the convergent 
expansions presented in~\cite{p}, which expresses the propagator $U(t)$ 
in terms of its {\it Floquet form}.  

Let us first begin with some conventions. We suppose that the interaction 
function $f$ is periodic with period $T_\omega = 2\pi/\omega$ and that its 
Fourier decomposition $f = \sum_n F_n e^{i n \omega t}$ has only a finite 
number of terms. Since $f$ is real, excluding the constant
term $F_0$, $f$ must have an even number of non-vanishing Fourier
coefficients, say $2J$, with $J \geq 1$. Denoting the set of integers  
$\{ n \in \Z, n \neq 0 \: | \: F_n \neq 0\}$ by $\{n_1, \ldots, n_{2J}\}$,
we may write
\begin{equation}
f(t) = F_0 + \sum_{a=1}^{2J} f_a e^{i n_a \omega t} \, , \nonumber
\end{equation}
with the convention that $n_a = - n_{2J - a+1}$, for all $1 \leq a
\leq J$, and with $f_a \equiv F_{n_a}$. Clearly $\overline{f_a} =
f_{2J -a +1}$, $1\leq a \leq J$. 

\subsection{Interactions with $F_0 = 0$}

In this case, the Fourier decomposition of the functions $q$ and $q^2$ can
be written as~\cite{qp,p}
\begin{equation}
q(t) = \sum_{m \in \Z} Q_m e^{i m \omega t} 
\quad \mbox{and} \quad 
q(t)^2 = \sum_{m \in \Z} Q_m^{(2)} e^{i m \omega t} \, . \nonumber
\end{equation}
The coefficients $Q_m$ and $Q_m^{(2)}$ are of basic importance in the
numerical computation of the propagator (\ref{eq:prop}). They can be
obtained in a closed form from the coefficients $F_m$ of $f$. Explicitly, 
we have~\cite{qp,p}   
\begin{equation} \label{eq:QQ}
Q_{m} = e^{i\gamma_f} \sum_{p_1 , \, \ldots , \, p_{2J} = 0}^{\infty} 
\delta(P, m) \prod_{a=1}^{2J} \ \left[
\frac{1}{p_a!} \left(\frac{f_a}{ n_a  \omega } \right)^{p_a}
\right] \, , 
\end{equation}
for all $m$, where
\begin{equation}
P \equiv P(p_1, \ldots , p_{2J}, n_1 , \ldots, n_{2J} ) := 
\sum_{b=1}^{2J}p_b n_b \, , \nonumber
\end{equation}
and
\begin{equation} 
\gamma_f := i\sum_{a=1}^{2J}
\frac{f_a}{n_a \omega} \, . \nonumber
\end{equation}
The symbol $\delta(P, m)$ denotes the Kr\"onecker delta:
$\delta(P, m) = 1$, if $P = m$ and $\delta(P, m) = 0$, otherwise.
To compute the Fourier coefficients of $q^2$, we simply note that $q^2$ is
obtained from $q$ by the substitution $f \to 2f$. Hence, for all $m$,  
\begin{equation} \label{eq:QQ2} 
Q_{m}^{(2)}  =  e^{2i\gamma_f} \sum_{p_1 , \, \ldots , \,
p_{2J} = 0 }^{\infty} \delta\left(P, m \right) \prod_{a=1}^{2J}
\ \left[ \frac{1}{p_a !}  \left( \frac{2f_a}{ n_a  \omega }
\right)^{p_a} \right]  \, .
\end{equation}
Formulas (\ref{eq:QQ}) and (\ref{eq:QQ2}) can be computed either 
analytically (in some cases) or numerically. For the monochromatic 
interactions, a closed form in terms of Bessel functions of first kind is 
obtained. The exact result will be presented in Section~\ref{sec:results}.

Once we know the coefficients $Q_m$ and $Q_m^{(2)}$, we proceed to compute
the particular solution (\ref{eq:Ansatz}) or (\ref{eq:AnsatzII}) of the
generalised Riccati equation (\ref{eq:Riccati}). To decide whether we use
(\ref{eq:Ansatz}) or (\ref{eq:AnsatzII}), we must check which 
condition I, II or III is satisfied by $f$. Hence, we have to look at the 
mean values of $\calQ_0$, $\calQ_1$ and $\calQ_3$. One obviously has
$
M(\calQ_0) = M(q^2) = Q_0^{(2)} 
$.
To obtain $M(\calQ_1)$ and $M(\calQ_3)$, we need the Fourier decompositions
of $\calQ_1$ and $\calQ_3$. From the definitions (\ref{eq:Q1}) and
(\ref{eq:Q3}), after some simple computations, we get
\begin{equation}
M(\calQ_1) = \frac{i}{\omega} \sum_{m \in \Z \atop m \neq 0}
\frac{\overline{Q_{-m}^{(2)}} \left(Q_0^{(2)} - Q^{(2)}_{-m}\right)}{m} \nonumber
\end{equation} 
and
\begin{eqnarray}
M(\calQ_3) &=& -\frac{1}{\omega^2} 
\sum_{n, m \in \Z \atop n \neq 0 , m \neq 0 }
\frac{\overline{Q_{-m}^{(2)}}}{n \, m} \left(Q_0^{(2)} Q_n^{(2)} \right. 
\nonumber \\
&-& \left. Q_0^{(2)} Q_{n-m}^{(2)} 
+ Q_{-n}^{(2)} Q_{n-m}^{(2)} \right) \, . \nonumber
\end{eqnarray}
The numerical value of $M(\calQ_1)$ and $M(\calQ_3)$ can be calculated
trivially from the above expressions once $Q_m^{(2)}$ are known. 

\subsubsection{Computing $g(t)$ in Cases I and III}

We remember that condition I applies whenever $f$ satisfies 
$M(\calQ_0) \neq 0$. In this case, by properly fixing the constants 
$\alpha_n$, we can completely eliminate the
secular terms from the functions $c_n$~\cite{qp,p}. 
For this reason, we may write 
\begin{equation}
c_n(t) = \sum_{m \in \Z} C_m^{(n)} e^{i m \omega t} \, . \nonumber
\end{equation}
The Fourier coefficients $C_m^{(n)}$ are obtained from 
equations (\ref{eq:c1})-(\ref{eq:cn}). Their inductive structure is given
by the relations~\cite{qp}
\begin{eqnarray}
C_{m}^{(1)} & = & \alpha_1 Q_{m} \, , \nonumber \\
C_{m}^{(2)} & = & \sum_{n_1 \in \Z \atop n_1 \neq
0} \frac{ Q_{m - n_1}\left( \alpha_1^2 Q^{(2)}_{n_1} -
\overline{Q^{(2)}_{-n_1}}
\right)}{n_1 \omega} \, + \, \alpha_2 Q_{m } \, , \nonumber \\
C_{m}^{(n)} & = & \sum_{n_1, \, n_2 \in \Z \atop
n_1 + n_2 \neq 0} \frac{Q_{m - (n_1 + n_2)} }{(n_1 + n_2) \omega}
\left(\sum_{p=1}^{n-1} C_{n_1}^{(p)}C_{n_2}^{(n-p)} \right) 
\nonumber \\ &+& \alpha_n Q_{m} \, , \quad \text{for } n \geq 3 \, . 
\nonumber
\end{eqnarray}
The constants $\alpha_n$ have closed forms in terms of the coefficients
$Q_m^{(2)}$ and $C_m^{(p)}$, for $p \leq n-1$. Since they involve somewhat
large expressions, we refrain from writing them
here. The complete expressions can be found in~\cite{qp}. 
We may see that the whole inductive structure of the coefficients 
$C_m^{(n)}$ is known and, therefore, the computation of $c_n$ is just a 
matter of numerically evaluating the above expressions. 

Since in case I $g$ is given by (\ref{eq:Ansatz}), we may write
\begin{equation} \label{eq:gt}
g(t) = q(t) \sum_{n = 1}^{\infty} c_n(t) \epsilon^n =: \sum_{n=1}^{\infty}
\left(\sum_{m \in \Z} G^{(n)}_m e^{i m \omega t}\right) \epsilon^n \, ,
\end{equation} 
where $G_m^{(n)}$ is given by the convolution
\begin{equation} 
G^{(n)}_m = \sum_{p \in \Z} Q_{m - p} C_{p}^{(n)} \, , \nonumber
\end{equation}
whose numerical value can be easily computed since we already know $Q_m$
and $C_m^{(n)}$, for all $m$ and $n$. This gives $g$ for all $t$ under
condition I. For condition III ($M(\calQ_0) = M(\calQ_1) = 0$, but
$M(\calQ_3) \neq 0$) one has essentially the same solution except
for the constants $\alpha_n$ which are calculated differently from
condition I. Their formulas can be found in~\cite{dacjcab1}.

\subsubsection{Computing $g(t)$ in Case II}

Condition II applies whenever $f$ satisfies $M(\calQ_0) = 0$, but
$M(\calQ_1) \neq 0$. In this case, the perturbative solution of the
generalised Riccati equation (\ref{eq:Riccati}) is given by
(\ref{eq:AnsatzII}). The constants $\beta_n$ which appear in the functions
$e_n$ (equations (\ref{eq:e1})-(\ref{eq:en})) can be chosen such
that no secular terms emerge~\cite{qp}. Hence, with this particular
choice of the constants $\beta_n$, we may write
\begin{equation} 
e_n(t) = \sum_{m \in \Z} E_m^{(n)} e^{i m \omega t} \, . \nonumber
\end{equation}
The recursive structure of the coefficients $E_m^{(n)}$ is given
by~\cite{qp}
\begin{eqnarray}
E_m^{(1)} &=& - \sum_{n \in \Z \atop n \neq 0} \frac{Q_{m-n} 
\overline{Q_{-n}^{(2)}}}{n \omega} + Q_m \left(\beta_1 + \sum_{n \in \Z
\atop n \neq 0} \frac{\overline{Q_{-n}^{(2)}}}{n \omega}\right) \, ,
\nonumber \\
E_m^{(n)} &=& \sum_{n_1, n_2 \in \Z \atop n_1 + n_2 \neq 0} 
\frac{Q_{m-n_1-n_2}}{(n_1+n_2) \omega}
\left(\sum_{p=1}^{n-1} E^{(p)}_{n_1} E^{(n-p)}_{n_2} \right)
\nonumber \\
&-& Q_m \sum_{n_1, n_2 \in \Z \atop n_1 + n_2 \neq 0} 
\left(\sum_{p=1}^{n-1} E^{(p)}_{n_1} E^{(n-p)}_{n_2}\right) 
\frac{1}{(n_1+n_2)\omega} 
\nonumber \\
&+& \beta_n Q_m \, , \quad \text{for } n \geq 2 \, .
\nonumber
\end{eqnarray} 
The constants $\beta_n$ have also closed forms~\cite{qp} in terms of the 
known $Q_m^{(2)}$ and $E_m^{(p)}$, for all $m$ and $p \leq n - 1$.
Thus, the solution (\ref{eq:AnsatzII}) can be evaluated numerically in any
order for all $t$. For future convenience, we write
\begin{equation} \label{eq:gtII}
g(t) = q(t) \sum_{n = 1}^{\infty} e_n(t) \lambda^n =: \sum_{n=1}^{\infty}
\left(\sum_{m \in \Z} \widetilde{G}^{(n)}_m e^{i m \omega t}\right) 
\lambda^n \, ,
\end{equation} 
where $\widetilde{G}_m^{(n)}$ is given by
\begin{equation}
\widetilde{G}^{(n)}_m = \sum_{p \in \Z} Q_{m - p} E_{p}^{(n)} \, ,
\nonumber
\end{equation}
which completely specifies $g$ for all $t$ under condition II.

\subsubsection{Computing the Propagator in Cases I, II and III}
\label{sssec:prop}

Once the coefficients $G_m^{(n)}$ (for cases I and III) or 
$\widetilde{G}_m^{(n)}$ (for case II) are known, the propagator $U(t)$
expressed in (\ref{eq:prop}) can be computed in a straightforward manner, 
as will be shown now. We illustrate our procedure with the coefficients
$G_m^{(n)}$ of conditions I and III. For condition II, where the 
coefficients of $g$ are $\widetilde{G}_m^{(n)}$, the discussed procedure 
has to be adapted with self-evident modifications.

We begin by defining
\begin{equation} \label{eq:Gm}
G_m(\epsilon) := \sum_{n = 1}^{\infty} G_m^{(n)} \epsilon^n \, .
\end{equation}
The secular frequency (\ref{eq:secular}) is clearly given by (see
(\ref{eq:gt})) 
\begin{equation}
\Omega \equiv \Omega(\epsilon) = \sum_{n=1}^{\infty} 
G_0^{(n)} \epsilon^n \, ,
\nonumber
\end{equation}
since we are supposing that $F_0 = 0$. Hence, by (\ref{eq:gt}),
\begin{equation} \label{eq:gtgz}
g(t) = \Omega + \sum_{m \in \Z \atop m \neq 0} G_m(\epsilon) 
e^{i m \omega t} \, .
\end{equation}
Looking at expression (\ref{eq:prop}) for the propagator, we see that
the Fourier series of $R$ (see (\ref{eq:R})) can be computed if we
first find the Fourier decomposition of
\begin{equation} 
W(t) := \exp{\left(-i \int_0^t g(\tau) \, d\tau \right)} \, . \nonumber
\end{equation}
Indeed, since $R = \overline{q} \, W$, we obtain the Fourier
series of $R$ by taking a convolution of the coefficients 
of $\overline{q}$ and $W$.

It is easy to see that
\begin{equation} \label{eq:Wdet2} 
W(t) = e^{i \gamma_f(\epsilon)} \, e^{-i \Omega t} \,
\exp{\left(-\sum_{m \in \Z} H_m \, e^{i m \omega t} \right)} \, ,
\end{equation}
with
\begin{equation} 
H_m \equiv H_m(\epsilon) := \frac{G_m(\epsilon)}{m\omega} 
\, , \quad \text{for } m \neq 0 
\nonumber  
\end{equation}
and $H_0 = 0$. Moreover, 
\begin{equation}
\gamma_f(\epsilon) := i \sum_{m \in \Z} H_m \, .
\nonumber
\end{equation}
Writing 
\begin{equation} \label{eq:W1}
W(t) \; = \; e^{- i \Omega t} \sum_{m \in \Z} W_m \, e^{i m \omega t} \, ,
\end{equation}
and using (\ref{eq:Wdet2}), we find
\begin{eqnarray}
W_m \equiv W_m(\epsilon) = 
e^{-i \gamma_f(\epsilon)}
 \left(
       -H_m +  \sum_{p=1}^{\infty}\frac{(-1)^{p+1}}{(p+1)!}
 \right. \nonumber \\
 \times 
 \left.           
       \sum_{n_1 , \ldots ,\,  n_p \in \Z}
                              H_{n_1} \cdots H_{n_p} H_{m -N_p}
 \right) , \quad \text{for } m \neq 0  
\nonumber
\end{eqnarray}
and
\begin{eqnarray}
W_0 \equiv W_0(\epsilon) = 
e^{-i \gamma_f(\epsilon)}
 \left(
       1 +  \sum_{p=1}^{\infty}\frac{(-1)^{p+1}}{(p+1)!}
 \right. \nonumber \\
 \times 
 \left.           
       \sum_{n_1 , \ldots ,\,  n_p \in \Z}
                              H_{n_1} \cdots H_{n_p} H_{ -N_p}
 \right) ,
\nonumber
\end{eqnarray}
with $N_p := \sum_{a = 1}^p n_a$. Using now (\ref{eq:W1}) and the fact that
$R = \overline{q} \, W$, we conclude that $R$ can be written as
\begin{equation} \label{eq:Rt}
R(t) = e^{-i \Omega t } \sum_{m \in \Z} R_m \, 
e^{i m \omega t} \, ,
\end{equation}
with the coefficients $R_m$ given by the convolution
\begin{equation} \label{eq:Rm}
R_m = \sum_{p \in \Z} \overline{Q_{p-m}} W_p \, .
\end{equation}
This finishes with the computation of $R$ in terms of its Fourier series
(\ref{eq:Rt}). We note from formulas (\ref{eq:Gm})-(\ref{eq:Rm})
that once the coefficients $G_m^{(n)}$ of $g$ are given, we can numerically
evaluate $G_m$, $H_m$, $W_m$ and, hence, $R_m$ with a trivial computer
code.

Next we proceed to compute the Fourier series of $S$ (see (\ref{eq:St})).
First we find the Fourier coefficients of $R^{-2}$. This is an easy task
since $R^{-2}$ is obtained from $R$ by replacing $(f+g) \to
-2(f+g)$. Hence, we must replace $H_m \to -2H_m$ and $\overline{q} \to
q^2$. Consequently, we get   
\begin{equation} \label{eq:Rtm2}
R(t)^{-2} = e^{2 i \Omega t} \sum_{m \in \Z} R_{m}^{(-2)} \,
e^{i m \omega t} \, ,
\end{equation}
where
\begin{equation} 
R_m^{(-2)} = \sum_{p \in \Z} Q_{m-p}^{(2)} W_p^{(-2)} \, .
\nonumber
\end{equation}
with
\begin{eqnarray}
W_m^{(-2)} \equiv W_m^{(-2)}(\epsilon) = 
e^{2 i \gamma_f(\epsilon)}
 \left(
       2H_m +  \sum_{p=1}^{\infty}\frac{2^{p+1}}{(p+1)!}
 \right. \nonumber \\
 \times 
 \left.           
       \sum_{n_1 , \ldots ,\,  n_p \in \Z}
                              H_{n_1} \cdots H_{n_p} H_{m -N_p}
 \right) , \quad \text{for } m \neq 0  
\nonumber
\end{eqnarray}
and
\begin{eqnarray}
W_0^{(-2)} \equiv W_0^{(-2)}(\epsilon) = 
e^{2i  \gamma_f(\epsilon)}
 \left(
       1 +  \sum_{p=1}^{\infty}\frac{2^{p+1}}{(p+1)!}
 \right. \nonumber \\
 \times 
 \left.           
       \sum_{n_1 , \ldots ,\,  n_p \in \Z}
                              H_{n_1} \cdots H_{n_p} H_{ -N_p}
 \right) .
\nonumber
\end{eqnarray}
Now $S$ is obtained by a simple integration of $R^{-2}$. A trivial
computation from (\ref{eq:Rtm2}), gives
\begin{equation} \label{eq:Stt}
S(t) =  \sigma_0 + e^{2 i \Omega t} \sum_{m \in \Z} S_m \,
e^{i m \omega t} \, ,
\end{equation}
with
\begin{equation} \label{eq:Sm}
S_m := -i \frac{R_m^{(-2)}}{m \omega + 2 \Omega}
\quad \text{and} \quad
\sigma_0:= - \sum_{m \in \Z} S_m \, .
\end{equation}
We assume that $m \omega + 2 \Omega \neq 0$ for all $m \in \Z$ (see the
discussion of {\it crossings} in~\cite{p}).

We have found expressions for $R$ and $S$ in terms of its Fourier series. 
This series converge absolutely and uniformly as we showed
in~\cite{p,dacjcab1}. To compute the propagator given in (\ref{eq:prop}),
we still need $g_0$, which can be easily  obtained from (\ref{eq:gtgz}):
\begin{equation} \label{eq:g0}
g_0 \equiv g(0) = \Omega + \sum_{m \in \Z \atop m \neq 0} G_m(\epsilon) \, .
\end{equation}
Formulas (\ref{eq:Rt}), (\ref{eq:Stt}) and (\ref{eq:g0}) can now be used to
evaluate $U(t)$ for all times. We stress that these formulas depend
essentially on the Fourier coefficients $G_m^{(n)}$ of $g$. These, in turn,
depend on $C_m^{(n)}$ which are direct linked to $Q_m$ and $Q_m^{(2)}$,
derived from the Fourier coefficients of the interaction function $f$. 
In short, to help us visualise the necessary steps towards the computation 
of $U(t)$, we may draw the following ``chain'':
\begin{equation}
\begin{array}{ccccccc}
f(t)            & \to  & F_m      & \to  & Q_m, Q_m^{(2)} & \to  & \alpha_n\vs \\
                &      &          &      &                &      & \vs\dto     \\
H_m             & \lto & \Omega   & \lto & G_m^{(n)}, g_0 & \lto & C_m^{(n)}\vs\\
\dto            &      &          &      &                &      & \vs         \\
W_m, W_m^{(-2)} & \to  & R_m, S_m & \to  & R(t), S(t)     & \to  & U(t) \, .
\end{array}
\nonumber
\end{equation}

\subsection{Interactions with $F_0 \neq 0$} \label{ssec:fneq0}

When $F_0 \neq 0$, one automatically has
$M(\calQ_0) = 0$, except perhaps when $2F_0 = k \omega$, for some integer
$k$. These facts were shown in~\cite{qp,p}. Since the situation where $2 F_0
= k \omega$, for some integer $k$, was nowhere investigated in our previous
works~\cite{qp,p,dacjcab1}, we will ignore this possibility by henceforth 
assuming that $2F_0 \neq k \omega$, for all $k \in \Z$. Since 
$M(\calQ_0) = 0$, condition I is never satisfied when $F_0 \neq
0$. The convergence of our expansions for $F_0 \neq 0$ in condition III,
however, has not yet been studied. Hence, we will only consider condition
II. In this case~\cite{p}, the function $q$ turns to be
\begin{equation} \label{eq:newq}
q(t) = e^{i F_0 t} \sum_{m \in \Z} Q_m e^{i m \omega t}
\end{equation}
and the functions $e_n$ (see (\ref{eq:e1})-(\ref{eq:en})),
\begin{equation} \label{eq:newen}
e_n(t) = e^{-i F_0 t} \sum_{m \in \Z} E^{(n)}_m e^{i m \omega t} \, .
\end{equation}
When $F_0 \neq 0$, the coefficients $E_m^{(n)}$ in the above
expression assume a simple form~\cite{p}:
\begin{eqnarray}
E_m^{(1)} &=& \sum_{a \in \Z} \frac{Q_{m+a}
\overline{Q_{a}^{(2)}}}{a \omega + 2F_0} \, ,
\nonumber \\
E_m^{(n)} &=& \sum_{p=1}^{n-1} \sum_{a, b \in \Z} \frac{Q_{m-a-b}
E_{a}^{(p)} E_{b}^{(n-p)}}{(a+b) \omega - 2F_0} \, , \quad \text{for } n
\geq 2 \, . \nonumber
\end{eqnarray}    
Finally, from (\ref{eq:newq}) and (\ref{eq:newen}), we conclude that $g$
given in (\ref{eq:gtII}) assumes the form
\begin{equation} 
g(t) = \sum_{m \in \Z} \widetilde{G}_m e^{i m \omega t} \, ,   
\nonumber
\end{equation}  
where
\begin{equation} 
\widetilde{G}_m \equiv \widetilde{G}_m(\epsilon) = \sum_{n=1}^{\infty}
\widetilde{G}_m^{(n)} \lambda^n \, , 
\nonumber
\end{equation} 
with $\lambda = \epsilon^2$ and
\begin{equation}
\widetilde{G}_m^{(n)} = \sum_{p \in \Z} Q_{m - p} E^{(n)}_p \, .
\nonumber
\end{equation}
This gives $g$ in terms of its Fourier series. 

To compute the propagator $U(t)$ accordantly to (\ref{eq:prop}), all
we have to do is follow the procedure detailed in
Section~\ref{sssec:prop} with the replacement of $G_m^{(n)} \to
\widetilde{G}_m^{(n)}$ and of $\epsilon \to \lambda$. 


\section{Numerical Results for the Monochromatic Interactions} 
\label{sec:results}

We now apply the algorithm discussed in Section~\ref{sec:numerical} to
study the system (\ref{eq:H1})-(\ref{eq:H2}) under the influence of
monochromatic interactions (ac-dc field):
\begin{equation} \label{eq:mono}
f(t) = F_0 + \varphi \cos(\omega t) \, ,
\end{equation}
which are relevance for many physical applications. With the
conventions introduced in Section~\ref{sec:numerical}, we have $J = 1$, $f_1
= f_2 = \varphi /2$, $n_1 = -n_2 = -1$. A simple application of formula
(\ref{eq:QQ2}) gives
\begin{equation}
\calQ_0(t) = \sum_{n \in \Z} J_n (\chi_1) e^{i (n + \chi_2) \omega t } \, ,
\nonumber
\end{equation}
where $J_n$ is the Bessel function of first kind and order $n$ and where we
defined $\chi_1 := 2 \varphi /\omega$ and $\chi_2 := 2F_0/\omega$.
Depending on the parameters $\chi_1$ and $\chi_2$, the function $f$
given in (\ref{eq:mono}) satisfies one of conditions I, II or III, except,
perhaps, for spurious situations. The detailed analysis of these facts can 
be found in~\cite{dacjcab1}. Table~\ref{tab:mono} summarises the conclusions
presented in~\cite{dacjcab1} and gives a classification of the conditions
satisfied by $f$ as a function of the parameters $\chi_1$ and $\chi_2$.

\begin{table}[h!]
\begin{center}
\begin{tabular}{cccc}
Label & $\chi_1$ & $\chi_2$ & Condition \\ \hline \\
\vspace{-0.65cm} \\
(A) & not a zero of $J_m$ & $-m \in \Z$ &  I   \\
(B) & zero of $J_m$       & $-m \in \Z$ &  III \\
(C) & any                 & not integer &  II  
\end{tabular}
\vskip0.15cm 
\caption{Classification of the conditions satisfied by the monochromatic
interactions $f(t) = F_0 + \varphi \cos(\omega t)$ as a function of the
parameters $\chi_1 = 2 \varphi /\omega$ and $\chi_2 = 2F_0/\omega$. We
labelled by (A), (B) and (C) the three possible cases.}
\label{tab:mono}
\end{center}
\end{table}
Although not indicated in Table~\ref{tab:mono} there is in case (C), 
on each interval $(k, k+1), k = 1, 2, \ldots$, a special value
$\chi_2^s$ of $\chi_2$ (depending on $\chi_1$) for which $M(\calQ_1) =
0$, and we would be out of condition II~\cite{dacjcab1}. We refrain
from studying this rather spurious situation here. A more detailed
analysis of this case can be found in~\cite{dacjcab1}.

We next show some graphical results of transition probabilities calculated
via the algorithm described in section~\ref{sec:numerical} for the
situations (A), (B) and (C) described in Table~\ref{tab:mono}. Let 
$\ket{\Phi_+} = \left(1 \atop 0\right)$ and 
$\ket{\Phi_-} = \left(0 \atop 1\right)$ be two orthogonal states of a
system described by (\ref{eq:H2}) (the eigenstates of the unperturbed
Hamiltonian $H_0$ of $H_1$ in (\ref{eq:H1})). The probability for the
transition from the initial state $\ket{\Phi_+}$ to the final state
$\ket{\Phi_-}$ at time $t$ is given by
\begin{equation} \label{eq:TP}
P(t) := |\bra{\Phi_+} U(t) \ket{\Phi_-}|^2 = |U_{12}(t)|^2 \, .
\end{equation}
We can evaluate $P(t)$ numerically using the methods described in
Section~\ref{sec:numerical} to compute $U(t)$. To estimate the accuracy of
our calculations, we tested the unitarity of the time evolution operator,
$U(t)^\dagger U(t) = \hat{1}$, and considered the quantity 
\begin{equation}
N(t) := |U_{11}(t)|^2 + |U_{12}(t)|^2 - 1 \, ,
\nonumber
\end{equation}
which should be identically equal to 0 for unitary $U(t)$.

\begin{figure}[b]
\centering{\epsfig{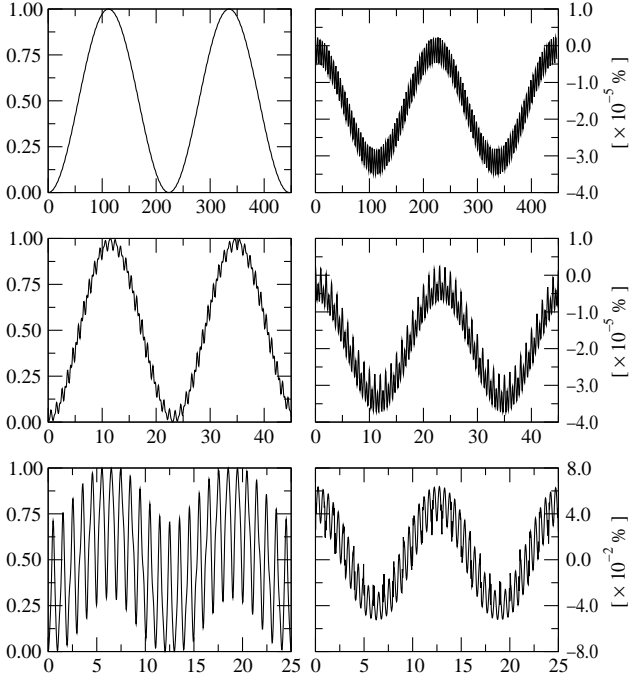}
\vskip0.25cm 
\caption{Plots of the transition probability $P(t)$ (left column) and
the deviation from unitarity $N(t)$ (right column) as a function of
time (measured in units of $T_\omega$) for various~$\epsilon$. 
We considered case (A) of Table~\ref{tab:mono}, with
$\omega = 1.0$, $\chi_1 = 2/\omega$ and $\chi_2 = 0$. 
We used $\epsilon = 0.01, 0.10, 0.40$ in the top, middle and bottom
rows, respectively.} 
\label{fig:A}}
\end{figure}

Let us first consider case (A) of Table~\ref{tab:mono} with $\omega =
1.0$, $\chi_1 = 2 / \omega$ and $\chi_2 = F_0 = 0$. Figure~\ref{fig:A} shows
graphs of $P(t)$ and $N(t)$ for $\epsilon = 0.01$, $\epsilon = 0.10$ and
$\epsilon = 0.40$, plotted from $t=0$ to $t = T_\Omega = 2\pi/\Omega$, in
units of $T_\omega = 2\pi /\omega$, the cycle of the external field. 
The calculations were performed using an expansion for $g$ up to 
$\calO(\epsilon^{25})$. We took all the Fourier coefficients
(generically called $\calF_m$) involved in the computations of $U(t)$ 
within the range $m = -40, \ldots, 40$. From the deviations of $N(t)$ from
0, we can infer very small errors in the calculations, leading to very
accurate values of $P(t)$. For $\epsilon = 0.01$ and $\epsilon =
0.10$, we have errors of the order of only $4.0 \times 10^{-5} \; \%$. For
$\epsilon = 0.40$, the errors jump to $6.0 \times 10^{-2}
\; \%$, indicating that the parameter $\epsilon$ is coming close to the
radius of convergence of our expansions. Since in case (A) we are
under condition I, the secular frequency $\Omega$ is given by
(\ref{eq:OmegaI}), hence $\Omega = \calO(\epsilon)$ (we choose $F_0 =
\chi_2 = 0$). For $\epsilon = 0.01$, from the full expansion of
$\Omega$ (see (\ref{eq:secular})), we get $T_\Omega \cong 450 
T_\omega$. For $\epsilon = 0.10$, $T_\Omega \cong 45 T_\omega$ and for
$\epsilon = 0.40$, $T_\Omega \cong 25 T_\omega$. We may notice from the
graphs of Figure~\ref{fig:A}, that the transition probability
$P(t)$ behaves like a Rabi oscillation with a
frequency $\Omega$, i.e. $P(t) \cong \sin^2(\Omega t)$. Small
oscillations of frequency $\omega$, with amplitudes of order
$|\epsilon|$, are superposed with this Rabi oscillation, leading to a
quasi-periodic evolution for the system (as ensured by the Floquet
theorem).

We now consider case (B) of Table~\ref{tab:mono} with $\omega = 10.0$,
$\chi_1 = x_1$, $x_1$ being the first positive zero of $J_0$, and
$\chi_2 = F_0 = 0$. Since we are under condition III and $F_0 = 0$ the
secular frequency becomes, according to (\ref{eq:OmegaIII}), $\Omega =
\calO(\epsilon^3)$, a fact first pointed in~\cite{bw}. This weak
dependence on $\epsilon$ indicates long transition times for the
probability amplitude $P(t)$. This phenomenon is known as the
dynamical localisation effect \cite{grossmann} (see \cite{sacchetti}
for a general criterion and for a more complete list of references).
To compute the solution in this case, $g$ was expanded up to
$\epsilon^6$. This contrasts with the expansion for $g$ in the
previous situation, where $g$ has been computed up to
$\calO(\epsilon^{25})$. The reason for this lies in the fact that
under condition III the expressions for the constants $\alpha_n$ (see
(\ref{eq:c1})-(\ref{eq:cn})) are somewhat more intricate than those of
condition I.  Fortunately, however, the expressions for the constants
$\alpha_1$, $\alpha_2$ and $\alpha_3$ are not so complicated and this
suffices to give an expansion of $g$ free of secular terms up to the
$\epsilon^6$ term (see~\cite{dacjcab1} for details). If we limit
ourself to study situations of small $\epsilon$, i.e. well inside of
the radius of convergence of our expansions, excellent results can be
achieved. To improve the accuracy of our results, it is better to use
a larger value of $\omega$.  

\begin{figure}[b]
\centering{\epsfig{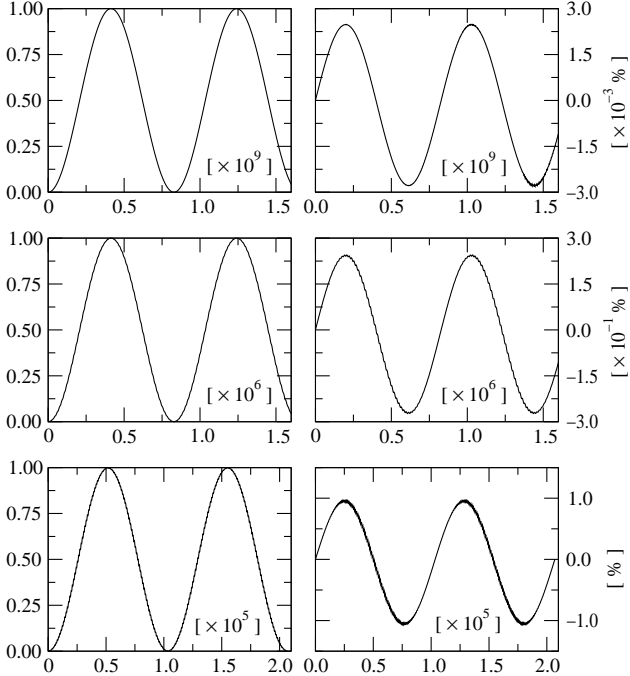}
\vskip0.25cm 
\caption{
The same of Figure~\ref{fig:A}, but we now considered case (B) of
Table~\ref{tab:mono}, with $\omega = 10.0$, $\chi_1 = x_1$, $x_1$
being the first positive zero of $J_0$ and $\chi_2 = 0$. We used 
$\epsilon = 0.01, 0.10, 0.20$ in the top, middle and bottom rows,
respectively. Note that the time scale is multiplied by the factor in
the square bracket.
}
\label{fig:B}}
\end{figure}

Figure~\ref{fig:B} shows graphs of the numerical values of $P(t)$ and
$N(t)$ for $\epsilon = 0.01$, $\epsilon = 0.10$ and $\epsilon = 0.20$,
plotted from $t=0$ to $t = T_\Omega = 2\pi/\Omega$, in units of
$T_\omega = 2\pi /\omega$. The errors in Figure~\ref{fig:B} (measured
out of the deviations of $N(t)$ from zero) are greater
than those presented in Figure~\ref{fig:A} due to the lower order
$\epsilon$-expansion of $g$ in the former case. However, for small
$\epsilon$, the errors are still small, being of the order of only
$3.0 \times 10^{-3} \; \%$ for $\epsilon = 0.01$ and of the order of
$3.0 \times 10^{-1} \; \%$ for $\epsilon = 0.10$. For larger
$\epsilon$ ($= 0.20$), the errors reach the value of about $1.0 \;
\%$, indicating that a higher order $\epsilon$-expansion is needed to
improve accuracy. We observe that the plots of $P(t)$ in
Figure~\ref{fig:B} have the same qualitative aspect of those in
Figure~\ref{fig:A}. There is a predominant Rabi oscillation (with
frequency $\Omega$) aspect superposed by minor oscillations of
frequency $\omega$ whose amplitudes are possible bounded by
$|\epsilon|^3$ terms. This fact, however, has yet not been proven
directly from our expansions, so it has to be faced more as a
qualitative analysis rather than a rigorous quantitative one. 

The main distinction between the graphs of Figure~\ref{fig:A} and
those of Figure~\ref{fig:B} is, undoubtedly, the Rabi oscillation
period $T_\Omega$. In the later, the long time needed for the system
to transit from the initial state $\ket{\Phi_+}$ to the final state
$\ket{\Phi_-}$, compared to the basic cycle $T_\omega$ of the external
perturbation, is the (approximate) dynamical localisation effect. The
secular period $T_\Omega$ obtained for the situations studied in
Figure~\ref{fig:B} were $T_\Omega \cong 1.6 \times 10^9 \, T_\omega$
for $\epsilon = 0.01$, $T_\Omega \cong 1.6 \times 10^6 \, T_\omega$
for $\epsilon = 0.10$ and $T_\Omega \cong 2.2 \times 10^5 \, T_\omega$
for $\epsilon = 0.20$. These are somewhat much larger values of
$T_\Omega$ than those of Figure~\ref{fig:A}.

\begin{figure}[b]
\centering{\epsfig{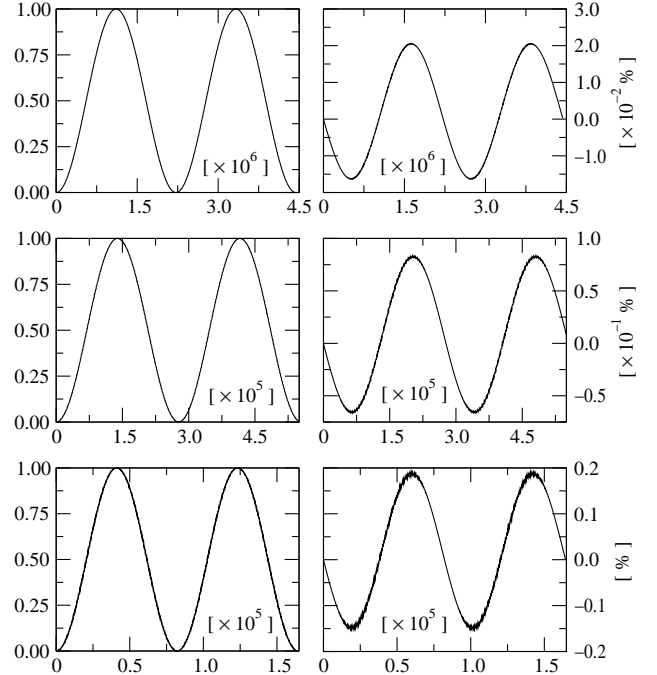}
\vskip0.25cm 
\caption{
The same of Figure~\ref{fig:B}, but now with $\chi_1 = x_2$, $x_2$ being
the second positive zero of $J_0$ and $\chi_2 = 0$. We used 
$\epsilon = 0.10, 0.20, 0.30$ in the top, middle and bottom rows,
respectively.
}
\label{fig:B2}}
\end{figure}

As we have mentioned, the graphs of Figure~\ref{fig:B} were computed
with $\chi_1 = x_1$, where $x_1$ is the first positive zero of $J_0$. It is
interesting to test our solution considering other possible zeros of
$J_0$. In Figure~\ref{fig:B2} we show plots of $P(t)$ and $N(t)$ calculated
with $\chi_2 = x_2$, the second positive zero of $J_0$ (we also took
$\omega = 10.0$ and, of course, $\chi_2 = 0$). The three situations
presented in Figure~\ref{fig:B2} correspond to $\epsilon = 0.10$, $\epsilon
= 0.20$ and $\epsilon = 0.30$. We may note that the qualitative behaviour
of $P(t)$ presented in Figure~\ref{fig:B2} is the same of 
Figure~\ref{fig:B}. In particular, the effect of dynamical localisation
is preserved since we still have $\Omega = \calO(\epsilon^3)$ when
$\chi_1 = x_2$. We may note from the deviation of unitarity $N(t)$
shown in Figure~\ref{fig:B2} that the errors, compared with the ones
in Figure~\ref{fig:B} for the same values of $\epsilon$, are smaller
by a factor $\sim 1/10$. We may understand this fact as follows: since
$x_2 > x_1$, the strength $\varphi$ of the interaction presented in 
Figure~\ref{fig:B2} is greater than in Figure~\ref{fig:B}, hence, for
the same values of $\epsilon$, the effective perturbation $\epsilon /
\varphi$ is smaller in the former case, leading to a more precise
perturbative computation.  

\begin{figure}[b]
\centering{\epsfig{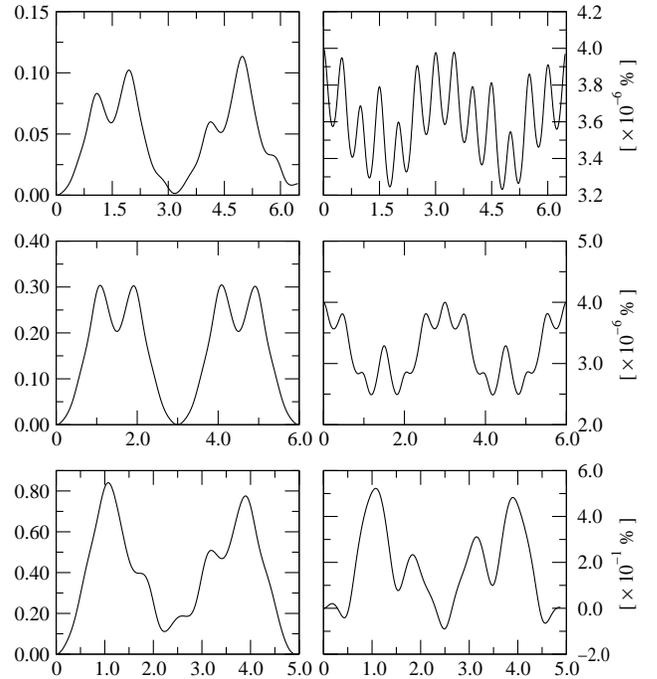}
\vskip0.25cm 
\caption{
The same of Figure~\ref{fig:A}, but we now considered case (C) of
Table~\ref{tab:mono}, with $\omega = 1.0$, $\chi_1 = 1.0$ and $\chi_2
= 0.3$ (not integer). We used $\epsilon = 0.05, 0.10, 0.20$ in the
top, middle and bottom rows, respectively.
}
\label{fig:C}}
\end{figure}

We now investigate case (C) of Table~\ref{tab:mono}. We consider
$\omega = 1.0$, $\chi_1 = 1.0$ and $\chi_2 = 0.3$. This implies
condition II and we have to follow the prescriptions of
Section~\ref{ssec:fneq0}.  Figure~\ref{fig:C} shows the results
obtained for $P(t)$ and $N(t)$ for three values of $\epsilon$: 0.05,
0.10 and 0.20.  The various plots were calculated using an expansion
of $g$ up to $\epsilon^{20}$. It was not necessary, thus, to use large
$\omega$ to ensure convergence of the expansions, as we did in the
situations presented in Figure~\ref{fig:B} (condition III). As usual,
$P(t)$ and $N(t)$ were computed from $t = 0$ to $t = T_\Omega$ in
units of $T_\omega$. The qualitative behaviour of the transition
probability is significantly different from the previous results
(Figures~\ref{fig:A} and~\ref{fig:B}). This is a consequence of the
non-vanishing constant field $F_0$ presented in the interaction $f$.
Indeed, the secular frequency $\Omega$ is now given by
(\ref{eq:OmegaII}) and, hence, is of order of $F_0$.  For $\epsilon =
0.05$, we obtained $T_\Omega \cong 6.3 T_\omega$, for $\epsilon =
0.15$, $T_\Omega \cong 6.0 T_\omega$ and for $\epsilon = 0.20$,
$T_\Omega \cong 5.0 T_\omega$. Since we choose $F_0$ as the same order
of $\omega$, there is a strong competition between the Rabi
oscillation (governed by $\Omega$) and the external field
oscillations. This leads to the patterns shown in Figure~\ref{fig:C},
which do not behave purely like $\sin^2(\Omega t)$. In particular,
$P(t) < 1$ for all times, leading to the conclusion that the state
$\ket{\Phi_+}$ never transits completely to $\ket{\Phi_-}$. One sees,
moreover, that the transition amplitude $P(t)$ tends to zero as
$\epsilon \to 0$ (c.p. (\ref{eq:Pdet}), below), much in contrast to
the cases pictured in Figs. \ref{fig:A}, \ref{fig:B} and \ref{fig:B2}.
The leading $\epsilon$-dependence of $P(t)$ can be algebraically
determined, in principle, but this was not yet performed due to the
complexity of our expansions.

It is interesting to note from the graphs of Figure~\ref{fig:C} that
the transition probability gets closer to~1 as $\epsilon$ increases.
It would seem that for $\epsilon$ large enough, we could have $P(t) > 1$.
However, our expansions would not converge in this case, since
$\epsilon$ would be greater than the radius of convergence. To
illustrate this situation, let us consider the trivial case where
$\chi_1 = 0$ and $\chi_2 = 2 F_0 /\omega$ is not an integer. Since now
$f(t) = F_0 = \omega \chi_2 /2$, the Schr\"odinger equation
(\ref{eq:H2}) becomes time-independent, with a simple Hamiltonian
given by
\begin{equation}
H_2 = \epsilon \sigma_1 + F_0 \sigma_3 = 
\left(
\begin{array}{cc}
F_0 & \epsilon \\
\epsilon      & -F_0
\end{array}
\right) \, .
\nonumber
\end{equation}
The propagator $U(t)$ can be computed by elementary methods
(f.i., by diagonalising the Hamiltonian), leading to
\begin{equation} \label{eq:Unp}
U(t) = \cos(\omega_0 t) \hat{1} 
-\frac{\sin(\omega_0 t)}{\omega_0}
\left( i F_0\sigma_3 + \epsilon \sigma_1\right) \, ,
\end{equation} 
where $\omega_0 := \sqrt{F_0^2 + \epsilon^2}$. Hence,
\begin{equation} \label{eq:Pdet}
P(t) = \frac{\epsilon^2}{F_0^2 + \epsilon^2} \sin^2(\omega_0 t) \, .
\end{equation}
Note that $P(t) < 1$ for all $\epsilon$ and all $F_0$.
For $f(t) = F_0$, the generalised Riccati equation (\ref{eq:Riccati})
admits a particular solution given by the constant 
$
g_0 = - F_0 + \sqrt{F_0^2 + \epsilon^2} 
$.
It was shown in~\cite{qp} that the $\epsilon$-expansion
(\ref{eq:AnsatzII}) for $f(t) = F_0$ coincides, as expected, with the
Taylor expansion (centred at $\epsilon = 0$) of $g_0$. Thus, our
method has a clearly restricted region of convergence defined by
$|\epsilon| < |F_0|$.

\begin{figure}[b]
\centering{\epsfig{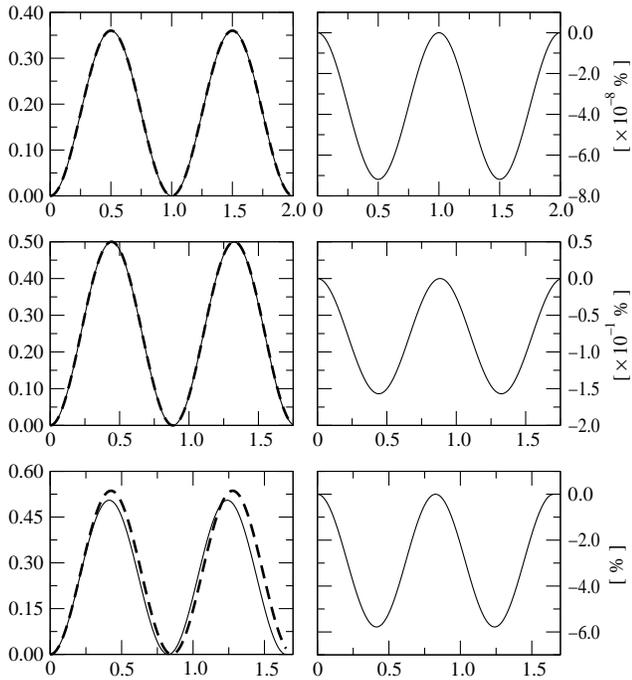}
\vskip0.25cm 
\caption{
The same of Figure~\ref{fig:C}, but now with $\chi_1 = 0$ and $\chi_2 =
0.8$ ($F_0 = 0.4$). We used $\epsilon = 0.30, 0.40, 0.43$ in the
top, middle and bottom rows, respectively. The dashed line represents
$P(t)$ calculated via formula (\ref{eq:Pdet}).
}
\label{fig:F0}}
\end{figure}

Let us see what these last considerations mean numerically. First we
set $F_0 = 0.4$.  Figure~\ref{fig:F0} shows plots of $P(t)$ and $N(t)$
for three critical values of $\epsilon$: $\epsilon = 0.30 < F_0$,
$\epsilon = 0.40 = F_0$ and $\epsilon = 0.43 > F_0$, where our
expansions are not supposed to converge. We also show in
Figure~\ref{fig:F0} (dashed lines) plots of $P(t)$ calculated
according to (\ref{eq:Pdet}). Looking at $N(t)$ and at the deviation
of the perturbatively computed transition probability from the one
calculated via (\ref{eq:Pdet}), we conclude that the region of
convergence of our expansions is restricted to $|\epsilon| < |F_0|$,
as expected. It is important to note that when $\epsilon < F_0$, our
computation of the propagator matches exactly with the
non-perturbative solution (\ref{eq:Unp}). This can be seen from
comparing $P(t)$ calculated perturbatively and non-perturbatively via
(\ref{eq:Pdet}), as shown in Figure~\ref{fig:F0}. 


\section{Final Remarks} \label{sec:conclusions}

We stress that, at least for the case of monochromatic interactions
examined in Section~\ref{sec:results}, the errors obtained are very
small and bounded as time increases (see the behaviour of $N(t)$ in
figures of Section~\ref{sec:results}). This is due to the absence of
secular terms in our perturbative expansions and its uniform
convergence in time. As a consequence, one can study the long-time
behaviour of the quantum system (\ref{eq:H1})-(\ref{eq:H2}) in a
controlled way.

Another important feature of our method is that it can be numerically
implemented with relatively simple computer codes. Indeed, the
algorithm described in Section~\ref{sec:numerical} to calculate the
unitary propagator $U(t)$ consists, essentially, in simple
computations involving the Fourier coefficients of the functions $q$
and $q^2$, which are known in a closed form (see (\ref{eq:QQ}) and
(\ref{eq:QQ2})). This is an important advantage against other
perturbative approaches, based f.i. on the Dyson expansion (\ref{eq:Dyson})
which, in general, cannot be evaluated in a simple manner (not to
mention the fact that such expansion is not uniform convergent in
time, as we have stressed in the introduction of this paper). To sum
up, our method is not only mathematically rigorous, but also very
useful for practical purposes, where it can be applied with great
generality, leading to very accurate results.


\acknowledgments

We are grateful to A. Sacchetti for discussions.
J.~C.~A.~Barata was partially supported by CNPq. 
D.~A.~Cortez was supported by FAPESP.




\end{document}